\documentclass{article}

\usepackage{spconf,amsmath,graphicx}
\usepackage{graphics}
\usepackage{bm}
\usepackage{amsthm}
\usepackage{amssymb}
\usepackage{nicefrac}
\usepackage{comment}
\usepackage{amsfonts}
\usepackage{color}
\usepackage{cite}
\usepackage[]{algorithm2e}
\usepackage[utf8]{inputenc} 
\usepackage[T1]{fontenc}
\usepackage[nolist]{acronym}
\usepackage[export]{adjustbox}
\usepackage[belowskip=5pt,aboveskip=2pt]{caption}
\graphicspath{{figures/}}
\begin{acronym}[SPACEEEEEE]
	\acro{AWGN}{additive white Gaussian noise}
	\acro{AMP}{approximate message passing}
	\acro{BER}{bit error rate}
	\acro{BPSK}{binary phase shift keying}
	\acro{BP}{belief propagation}
	\acro{CF}{cycle-free}
	\acro{CPM}{circulant permutation matrices}
	\acro{CN}{check node}
	\acro{CRAN}{cloud - radio access network}
	\acro{CSA}{coded slotted ALOHA}
	\acro{CSCG}{circularly-symmetric complex Gaussian}
	\acro{CSI}{channel state information}
	\acro{CU}{central unit}
	\acro{ESD}{empirical spectral distribution}
	\acro{FBL}{finite-blocklength}
	\acro{FEC}{forward error correction}
	\acro{GAMP}{generalized appoximate message passing}
	\acro{GQ}{generalized quadrangle}
	\acro{H-GAMP}{hybrid generalized approximate message passing}
	\acro{i.i.d.}{identical and independent distributed}
	\acro{IoT}{Internet of Things}
	\acro{IRSA}{irregular repetition slotted ALOHA}
	\acro{IWSN}{industrial wireless sensor network}
	\acro{JSC}{joint source-channel}
	\acro{LCR}{low-code rate}
	\acro{LDCD}{Low-density code-domain}
	\acro{LDPC}{low-density parity check}
	\acro{LDS}{low-density spreading}
	\acro{LLR}{log-likelihood ratio}
	\acro{LR}{likelihood ratio}
	\acro{LSD}{limiting spectral distribution}
	\acro{LSL}{large system limit}
	\acro{MF}{matched filter}
	\acro{MnAC}{many-access channel}
	\acro{MPA}{message passing algorithm}
	\acro{MRC}{maximum ratio combining}
	\acro{MAC}{multiple-access channel}
	\acro{ML}{maximum likelihood}
	\acro{mMTC}{massive machine-type communications}
	\acro{MUD}{multi-user detector}
	\acro{NOMA}{non-orthogonal multiple access}
	\acro{OFDM}{orthogonal frequency division multiplex}
	\acro{pdf}{probability distribution function}
	\acro{pmf}{probability mass function}
	\acro{QC}{quasi-cyclic}
	\acro{RB}{resource block}
	\acro{RE}{resource element}
	\acro{RRH}{remote radio head}
	\acro{SNR}{signal-to-noise ratio}
	\acro{SCMA}{sparse coded multiple access}
	\acro{SSC}{separate source-channel}
	\acro{SUMF}{single user matched filter}
	\acro{TAKTILUS}{\textit{Taktiles Internet f\"ur sichere und zeitsensitive Anwendungen der Industrie- und Prozessautomation}}
	\acro{TBRA}{type-based random access}
	\acro{TBMA}{type-based multiple access}
	\acro{UE}{user equipment}
	\acro{VN}{variable node}
	\acro{gNB}{gNodeB}
\end{acronym}
\title{Joint Source-Channel Coding and Bayesian Message Passing Detection \\ for Grant-Free Radio Access in IoT}
\name{Johannes Dommel$^*$, Zoran Utkovski$^*$, S\l awomir Sta\'{n}czak$^{*+}$ and Osvaldo Simeone$^\dagger$}
\address{$^*$Dept. of Wireless Communications and Networks, Fraunhofer Heinrich Hertz Institute, Berlin \thanks{J. Dommel and Z.Utkovski have received funding from the German Federal Ministry of Education and Research (BMBF) under grant 16KIS081.}\\
        $^{*+} $Network Information Theory Group, Dept. of Telecommunication Systems, Technical University of Berlin\\
        $^\dagger$ King's Centre for Learning \& Information Processing (KCLIP), Dept. of Informatics, King's College London \thanks{O. Simeone has received funding from the  European  Research  Council  under  the  European  Union’s Horizon  2020  research  and  innovation  program  under  grant  725731. }}
\begin{document}
\maketitle
\begin{abstract}
Consider an Internet-of-Things (IoT) system that monitors a number of multi-valued events through multiple sensors sharing the same bandwidth. 
Each sensor measures data correlated to one or more events, and communicates to the fusion center at a base station using grant-free random access whenever the corresponding event is active.
The base station aims at detecting the active events, and, for each active event, to determine a scalar value describing each active event's state. 
A conventional solution based on Separate Source-Channel (SSC) coding would use a separate codebook for each sensor and decode the sensors' transmitted packets at the base station in order to subsequently carry out events' detection. 
In contrast, this paper considers a potentially more efficient solution based on Joint Source-Channel (JSC) coding via a non-orthogonal generalization of Type-Based Multiple Access (TBMA). 
Accordingly, all sensors measuring the same event share the same codebook (with non-orthogonal codewords), and the base station directly detects the events' values without first performing individual decoding for each sensor. 
A novel Bayesian message-passing detection scheme is developed for the proposed TBMA-based protocol, and its performance is compared to conventional solutions.
\end{abstract}
\section{Introduction}
\label{sec:intro}
\textit{Motivation:}
In the context of the \ac{IoT}, \ac{mMTC} refers to communication set-ups in which a large number of \ac{IoT} devices, such as sensors, communicate sporadically  with a single receiver by transmitting short messages.
\setlength{\belowcaptionskip}{-10pt}
\begin{figure}[t]
\centering
\includegraphics[clip, trim=1.2cm 1cm 1.2cm 0cm,width=1\columnwidth]{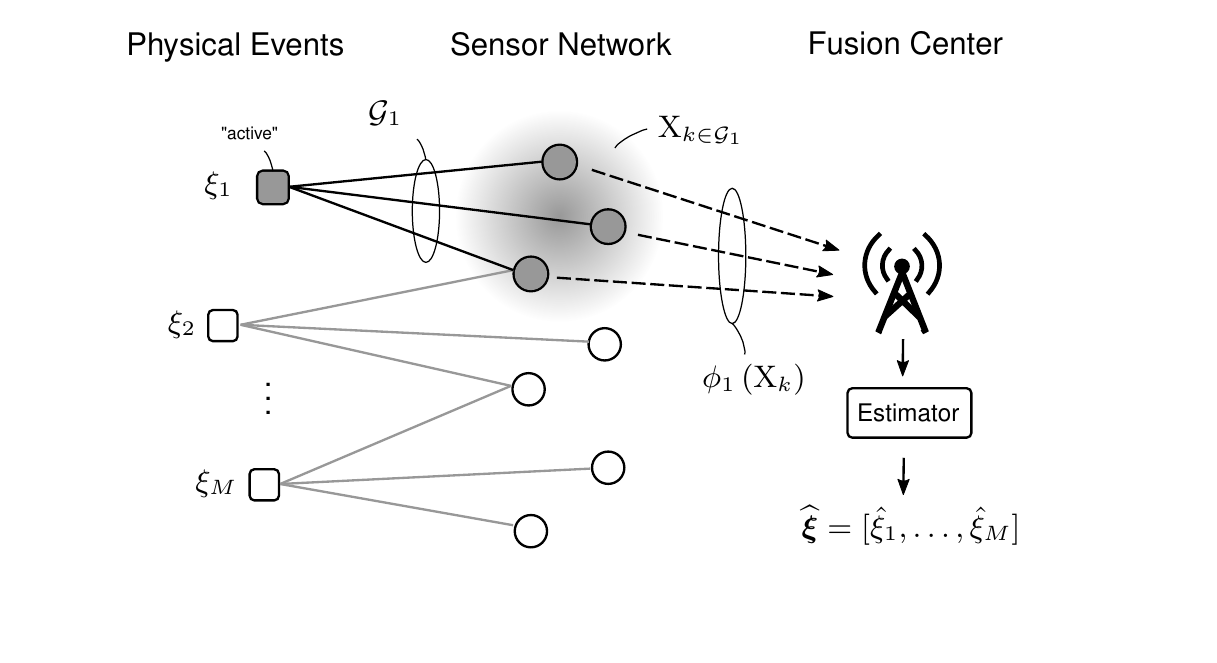}
\caption{Event-driven random access: A wireless sensor network monitors $M$ independent events, with each event $m$ being either inactive ($\xi_m=0$) or active with some associated scalar state value $\xi_m \in \{1,...,R\}$. Each sensor $k \in \mathcal{G}_m$ measuring a given event produces a local estimate  $\phi_m(X_k) \in \{0,1,...,R\}$ based on measurement $X_k$. Local estimates are conveyed to the fusion center for joint events' detection.}
\label{fig:event_based_scenario}
\end{figure}%
The design of mMTC communication protocols is challenged by the inadequacy of the conventional \ac{MAC} model that has provided the theoretical underpinning for the study of uplink transmission strategies. In fact, unlike the standard MAC model, mMTC systems are typically characterized by: (\emph{i}) small payloads \cite{polyanskiy10}; (\emph{ii}) uncoordinated access, possibly with grant-less or grant-free data transmission \cite{Shahab19}; (\emph{iii}) sparse user activity, with number of active users possibly exceeding the overall message blocklength \cite{chen14}; (\emph{iv}) correlated event-driven transmissions; and (\emph{v}) fusion-based decoding, whereby functions of multiple IoT sensors' measurements, rather than individual measurements, are of interest to the receiver. In this work, we propose a transmission protocol that accounts for all these aspects, allowing for grant-free uncoordinated short-packet transmissions of correlated measurements.\smallskip\\
\noindent\textit{Context and related work:} The problem of designing and analyzing mMTC protocols is currently being studied both from a fundamental information-theoretic standpoint and in terms of algorithms and protocols. Information-theoretic studies have focused mostly on the so-called many-user and unsourced random access models. In the many-user set-up, the number of users, which may be random, increases proportionally to the blocklength \cite{chen14}. This is in contrast to the standard large-system analysis of multiuser systems in which the blocklength is let to go to infinity before the number of users is made arbitrarily large \cite{shamai97}. In the unsourced random access model, users employ the same codebook, and the problem is identifying transmitted codewords, irrespective of the identity of the sender \cite{polyanskiy17}. Grant-free random access transmission protocols are under intense investigation, including in standardization bodies \cite{Shahab19}, with most work focusing on aspects (\emph{i})-(\emph{iii}) listed above (see \cite{Shahab19} for a review). 

Potential efficiency gains arising from consideration of correlated data generation (\emph{iv}) and fusion-based decoding (\emph{v}) have been addressed in past activity on sensor networks. Among the most notable results of this line of research, Type-Based Multiple Access (TBMA) is a transmission protocol that implements a form of Joint Source-Channel (JSC) coding, whereby each measurement value is assigned an orthogonal codeword and the receiver infers the parameter of interest from the histogram of the received measurements \cite{mergen06, Liu04, anandkumar07}. Recently, TBMA has been studied in a multi-cell set-up for IoT applications under centralized or decentralized decoding \cite{Kassab19}.\smallskip\\
%
%
%
%
%
%
%
\noindent\textit{Contributions:} In this paper, we propose a grant-free random access scheme that builds on JSC coding via a novel \emph{non-orthogonal} variant of TBMA. The method applies to a general multi-event setting. The non-orthogonality of codewords assigned to different measurement values makes standard decoders used in prior work on TBMA suboptimal and inefficient. To obviate this problem, we introduce a Bayesian Approximate Message Passing (AMP) detector\cite{rangan17} that jointly infers the events' activity and related scalar parameters (see also \cite{liu2018sparse} for a review of related uses of AMP in random access).\smallskip\\
\vspace{-10pt}
\section{Event-based Random Access for IoT:  System Model and Coding Schemes}
\label{sec:bayesian_framework}
\noindent\textit{Scenario and problem description:}
Consider the wireless \ac{IoT} network  illustrated in Fig.~\ref{fig:event_based_scenario},  consisting of $K$ devices tasked with monitoring $M$ ``events`` and transmitting event-related information to a centralized fusion center. Events $m=1,...,M$ are characterized by independent scalar random variable $\xi_m \in \{0, 1, \ldots, R\}$, with $P(\xi_m = 0)=1-\rho$ representing the probability that event $m$ is inactive. When the event is active, the event takes one of the values in the set $\{1,\ldots,R\}$, so that parameter $R$ measures the amount of information attached to the occurrence of an event.


%
%

%

%
%
%
%
Each device $k$ can simultaneously monitor a subset of events $\gamma(k)\subseteq\{1,...,M\}$. Therefore, the devices can be partitioned into $M$, generally overlapping, groups $\mathcal{G}_m = \{k\in\{1,...K\}: m \in \gamma(k)\}$, for $m = 1, \ldots, M$, with group $\mathcal{G}_m$ monitoring event $m$. Each device $k$ performs a local (real-valued) measurement $X_k$, which is generally correlated with all the variables $\xi_m$ for $m\in\gamma(k)$. For each event $m\in\gamma(k)$, the local measurement $X_k$ is mapped to a value $\phi_m(X_k)\in\{0,1, \ldots, R\}$, which represents a local estimate of event $m$. 
Upon evaluating the local estimates $\phi_m(X_k)$ for all $m\in\gamma(k)$, device $k$ maps each estimate $\phi_m(X_k)=r\in\{0,1, \ldots, R\}$ into a codeword $\boldsymbol{s}^m_{k,r}\in \mathbb{C}^{N \times 1}$ of length $N$ complex symbols. Note that we assume no knowledge of the channel coefficients at the devices. Codewords for each device $k$ and event $m$ are selected from a given codebook $\boldsymbol{S}^m_k = [\boldsymbol{s}^m_{k,0}, \ldots, \boldsymbol{s}^m_{k,R}]$ of generally non-orthogonal codewords (columns). For future reference, we also define the matrix $\boldsymbol{S}_k = [\boldsymbol{S}^1_k, \ldots, \boldsymbol{S}^M_k]$ that collects all codebooks of device $k$. All codewords are subject to a power constraint $ \|\boldsymbol{s}^m_{k,r}\|^2 \leq E$.\smallskip\\
\noindent \emph{Channel model:} We assume time synchronization and transmission over a block-fading channel model with coherence time–frequency span no smaller than that occupied by the codewords' duration. The signal received at the fusion center can hence be written as

\begin{equation}\label{eq:model1}
    \boldsymbol{y}=\sum_{k\in\mathcal{K}} h_k \sum_{m\in\gamma(k)} \boldsymbol{s}^m_{k,r} + \boldsymbol{w},
\end{equation}
where $h_k$ is the channel coefficient for the link between device $k$ and the fusion center, and $\boldsymbol{w}$ is \ac{i.i.d.} complex Gaussian noise with zero mean and variance $\sigma^2$. To simplify (\ref{eq:model1}), we define for each device $k$ the binary vector 
\begin{equation}
\boldsymbol{x}_k =[ (\boldsymbol{c}_k^1)^{T}, \ldots, (\boldsymbol{c}^M_k)^{T}]^{T} \in \{0, 1\}^{M(R+1) \times 1},
\label{eq:x}
\end{equation} 
where we have introduced the indicator vectors
\begin{align}
    \boldsymbol{c}_k^m = 
    \begin{cases}
    \boldsymbol{e}_{\phi_m(X_k)}   & \text{if}~m \in \gamma(k)\\
    \boldsymbol{e}_{0} & \text{otherwise}. \label{eq:x_k_input}
    \end{cases}
\end{align}
In (\ref{eq:x_k_input}) $\boldsymbol{e}_{r}$ is an $R+1$-dimensional binary vector with a single non-zero-entry at the $(r+1)$-th position. 
With this definition, the received signal can be described in matrix-notation as  
\begin{align}
    \boldsymbol{y} &= \sum_{k \in \mathcal{K}} h_k \boldsymbol{S}_k \boldsymbol{x}_k  + \boldsymbol{w}\label{eq:ssc_rx}\\
     &=\boldsymbol{A}\tilde{\boldsymbol{x}} + \boldsymbol{w},     
\end{align} 
with $\boldsymbol{A} = [\boldsymbol{S}_1, \ldots, \boldsymbol{S}_K]$ and $\tilde{\boldsymbol{x}} = [h_1\boldsymbol{x}_1^T, \ldots, h_K\boldsymbol{x}_K^T]^T$.\smallskip\\ 
%
%
%
%
%
%
\noindent \emph{Separate vs joint source-channel coding:} Depending on the structure of the codebooks, we will distinguish between two approaches. The  conventional approach based on separate source-channel coding (SSC) uses a distinct codebook $\bm{S}_k$ for each device $k$.  
In contrast, with joint source-channel coding (JSC), for any event $m$, all devices use a shared codebook $\boldsymbol{S}^m_k=\boldsymbol{S}^m= [\boldsymbol{s}^m_0, \ldots, \boldsymbol{s}^m_R]$ for all $k=1,\ldots,K$. As a result, all devices select the \textit{same} codeword $\boldsymbol{s}^m_{r}$ when producing the same local estimate $r$ for the $m$-th event. This approach can be considered as a generalization of
TBMA, given that the latter assumes a single event ($M=1$) and that the $R+1$ codewords in $\boldsymbol{S}^m$ are orthogonal. In the next section, we discuss receiver-side processing for both SSC and JSC. In all cases, we assume no channel state information at the receiver.
\section{Receiver Processing for Event-Based Random Access for IoT: A Bayesian Formulation}
\label{sec:decoding}
\begin{figure}[b]
\centering
\includegraphics[width=\columnwidth]{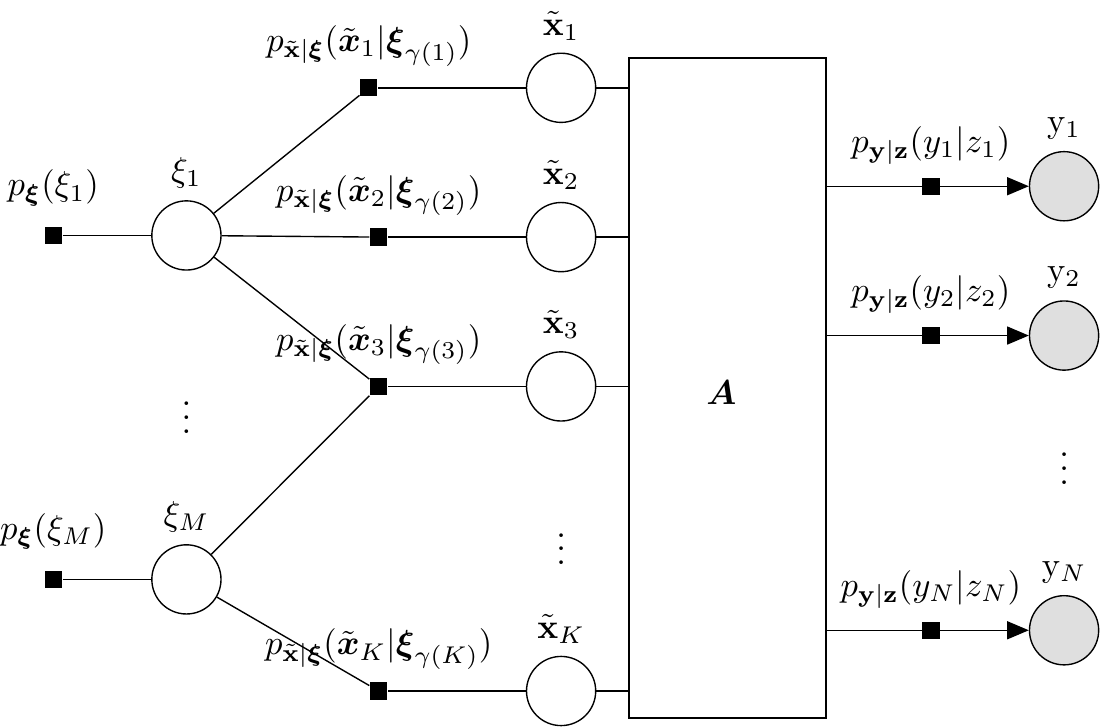}
\caption{Factor graph model representation of the relevant random variables for the design of the H-AMP detector.}
\label{fig:graphical_model}
\end{figure}

Standard decoding for TBMA is based on the assumption that the codewords assigned to the (single) event are orthogonal \cite{mergen06}. We propose here an AMP-based decoder that applies for any choice of the codebooks, including the choices discussed above for SSC and JSC.\smallskip\\ 
\noindent\textit{Graphical Model Representation:} To this end, we start by defining a factor graph that describes the joint distribution of the variables introduced above. Given the mentioned assumptions, the joint distribution $p_{\boldsymbol{\xi},\tilde{\mathbf{x}},\mathbf{y}}(\cdot,\cdot,\cdot)$ of the relevant variables $(\boldsymbol{\xi},\tilde{\boldsymbol{x}},\mathbf{y})$ factorizes as
\begin{align}
\prod_{m=1}^{M} p_{\boldsymbol{\xi}}(\xi_m) \prod_{k=1}^{K} p_{\tilde{\mathbf{x}}\vert \bm{\xi}}
(\tilde{\boldsymbol{x}}_k \vert \bm{\xi}_{\gamma(k)}) \prod_{i=1}^{N} p_{\mathbf{y}\vert \mathbf{z}}(y_i\vert z_i),  \label{eq:factorization}
\end{align}
where the second term, given the definition of vector $\tilde{\boldsymbol{x}}$, describes both the (deterministic) coding process and the fading channel statistics, and we have defined $z_i=\boldsymbol{a}_i^T\tilde{\boldsymbol{x}}$, with $\boldsymbol{a}_i$ being the $i$-th row of $\boldsymbol{A}$. The factor graph of the joint distribution (\ref{eq:factorization}) is illustrated in Fig.~\ref{fig:graphical_model}.
%
%
%
%
%
\smallskip\\
%
%
\noindent\textit{Bayesian Inference:} Given the factorization (\ref{eq:factorization}), the detector aims at computing the posterior distribution $p_{\bm{\xi}\vert \mathbf{y}}(\bm{\xi}\vert \boldsymbol{y})$ of the events' state vector $\bm{\xi}$ given the observation vector $\boldsymbol{y}$. While exact posterior inference is generally intractable, the Hybrid AMP (H-AMP) algorithm proposed in \cite{rangan17} provides an efficient solution with desirable empirical performance. H-AMP operates by iteratively exchanging soft information between two modules: the first carries out standard AMP by treating the entries of the vector $\tilde{\boldsymbol{x}}$ as independent, while the second refines the output of the first by leveraging the correlation structure of the entries of vector $\tilde{\boldsymbol{x}}$ in (\ref{eq:factorization}).

\section{Numerical Results}
\label{sec:num_res}
In this section, we perform numerical analysis to 
%
compare the performance of SSC and JSC in terms of decoding error as a function of the following key system parameters: the number  $M$ of observed events, the parameter $R$, the signature length $N$, the number $K$ of system devices, and the group $\{\mathcal{G}_m\}$ partitioning of the devices. 

%
%
Consider the special case in which each device observes only one of the $M=24$ events, with the same number of devices observing each event. This results in a partitioning of the device set into $M$ non-overlapping groups $\{\mathcal{G}_m\}$, each of size $G=K/M$. The channel coefficients $h_k$ are modeled as \ac{i.i.d.} circularly-symmetric complex Gaussian with zero mean and unit variance, $h_k\sim\mathcal{CN}(0,1)$. We recall that they are unknown to both devices and receiver. All codewords are generated randomly, with \ac{i.i.d.} circularly-symmetric complex Gaussian entries with zero-mean and variance $E/N$, for which the convergence of AMP has been studied rigorously~\cite{rangan17}. The average \ac{SNR} is defined as $\textrm{SNR} \doteq E/\sigma^2$. 
%
%
The figure of merit is the average decoding error rate, defined as
\begin{align}
    \text{Error Rate} &\doteq \frac{1}{M} \sum_{m = 1}^M \text{Pr}\left\{\xi_m \neq \hat{\xi}_m~\vert~\xi_m\right\}.
\end{align}

\begin{figure}[htb]
\begin{minipage}[b]{1.0\linewidth}
  \centering
  \centerline{\includegraphics[width=8cm]{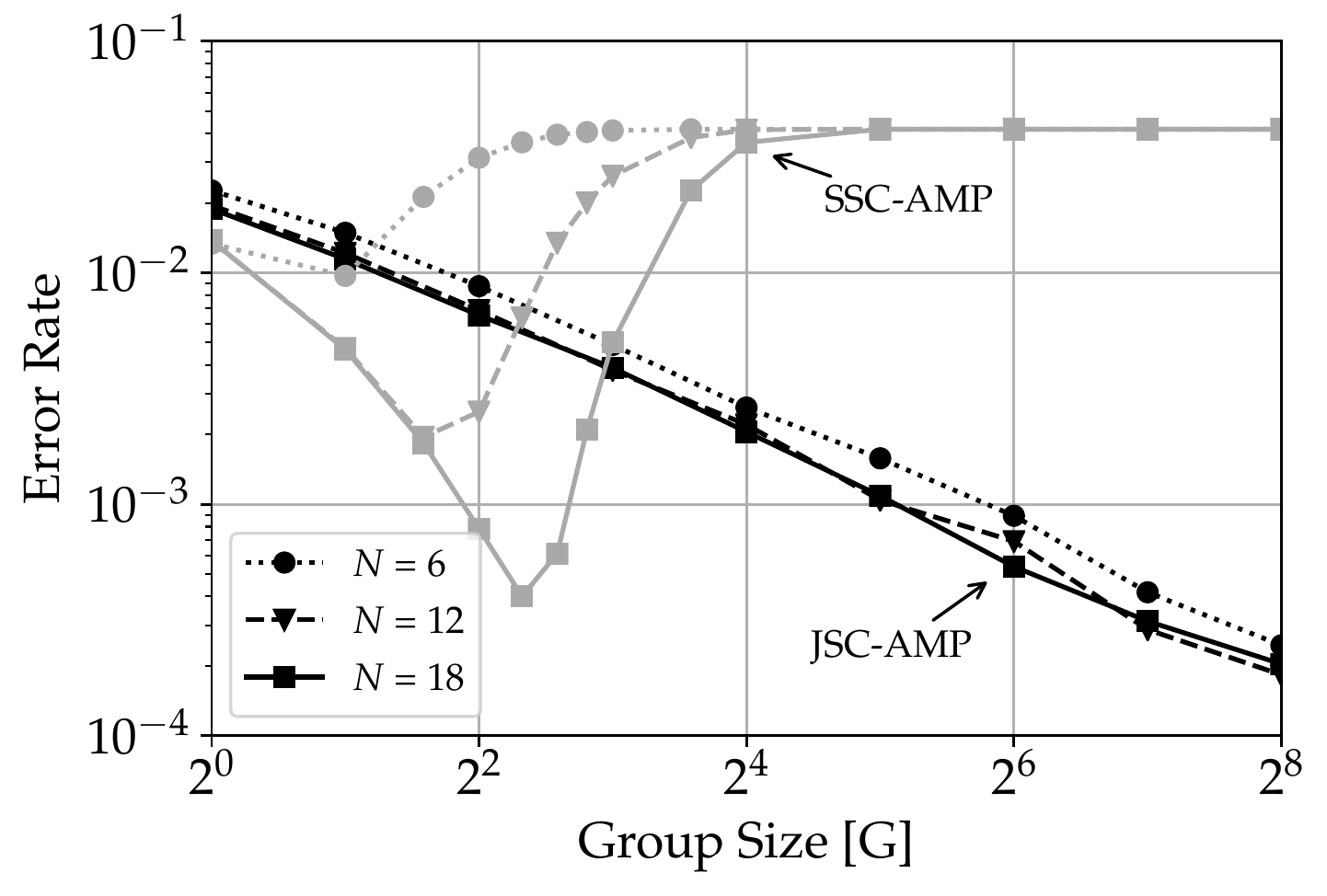}}
\end{minipage}
\caption{Decoding error-rate for SSC and JSC as function of the group size $G$, with the number $N$ of channel uses as parameter. The number of events is $M = 24$, the cardinality of the event-related variables is $R=1$, probability of activation $\rho = 0.1$ and $\text{SNR} = 12~\text{dB}$.} 
\label{fig:res_1}
\end{figure}

Fig.~\ref{fig:res_1} investigates the impact of the groups size $G$, i.e., of the number of devices configured to observe each of the $M$ events, on the decoding performance of SSC and JSC. We set $R=1$, so as to focus on the problem of events' activity detection. As $G$ increases, the total number of devices $K=GM$ also grows proportionally. For both SSC and JSC, the error rate is plotted as a function of $G$ for different values of $N$. 
It is observed SSC works well for small values of $G$, but it is significantly impaired as the group size $G$
increases beyond a given value. This is because SSC assigns different codewords to each device, and is hence able to support only a
maximum given number of devices in the group, for the given value of $N$, without causing a degradation of the error rate. In contrast, JSC is able to leverage larger group sizes to decrease the error rate below that achievable by SSC, particularly for small values of $N$. This is thanks to the fact that, with JSC, the signals sent by all devices producing the same local estimate of a given event are \emph{combined on air} through the choice of a common codebook. For example with $N=6$, SSC can achieve an error rate of at most $10^{-2}$
(for $G=2$), while JSC achieves an error rate of $2.6\cdot 10^{-3}$ for $G=16$, and $2.4\cdot 10^{-4}$ for $G=256$.

%
\setlength{\belowcaptionskip}{-10pt}
\begin{figure}[bth]
\begin{minipage}[b]{1.0\linewidth}
  \centering
  \centerline{\includegraphics[width=8cm]{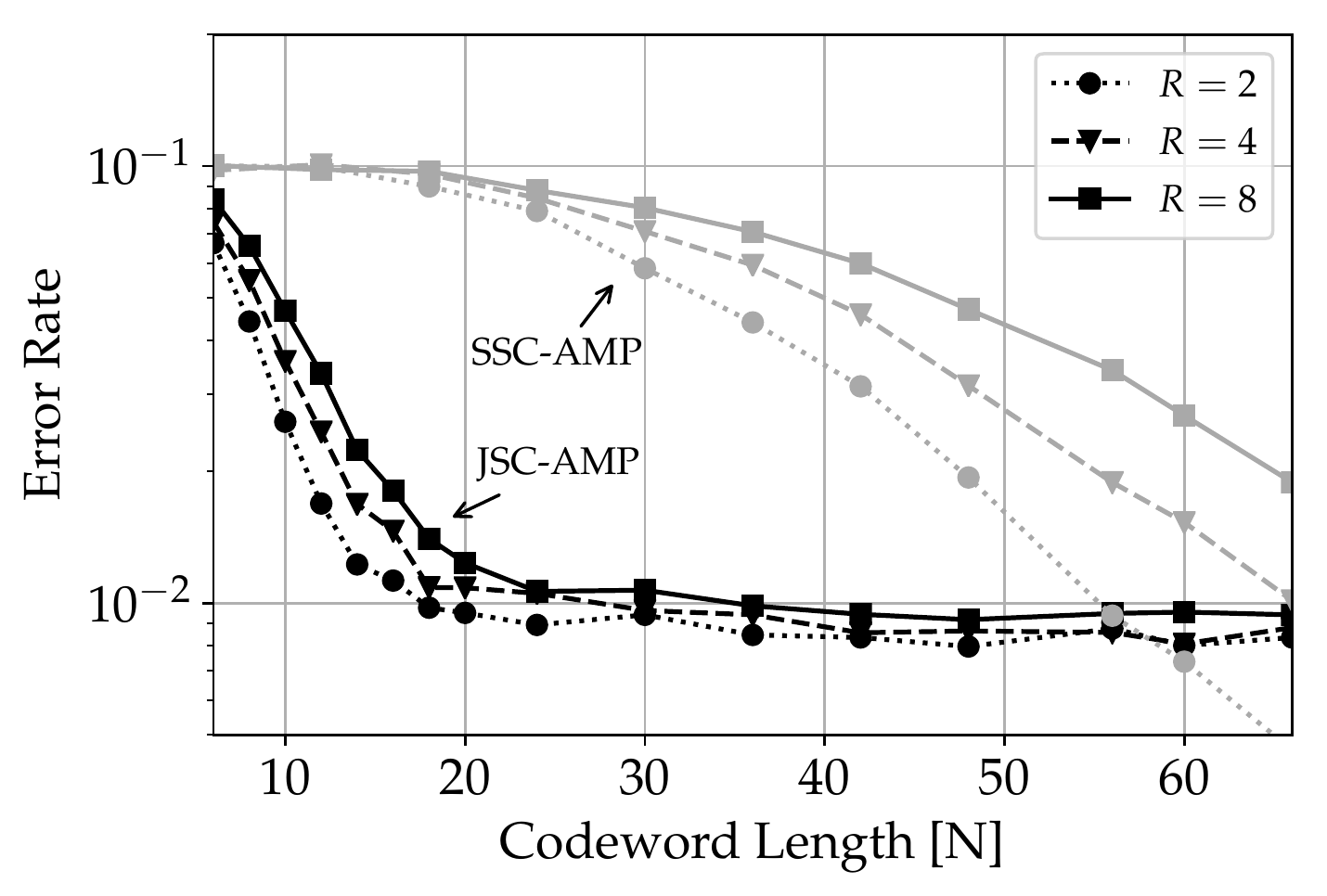}}
\end{minipage}
\caption{Decoding error-rate for SSC and JSC as function of the codewords' length $N$, with the cardinality of the event-related variables as parameter. The number of events is $M = 24$, the group size is $G=8$, the activation probability is $\rho = 0.1$ and $\text{SNR} = 12~\text{dB}$. }  
\label{fig:res_2}
\end{figure}
To gain more insight into this point, for the same set-up, in Fig.~\ref{fig:res_2}, we directly investigate the impact of the number $N$ of channel uses on the decoding error performance when $G = 8$. We vary here also the parameter $R$ in order to consider the more general case in which active events are characterized by a scalar quantity of interest (of cardinality $R$). Here, we consider the non-zero values of each event to be uniformly distributed with probability $\rho/R$. 
In line with the discussion above, we observe that JSC can vastly outperform SSC for small values of $N$, here up to around $N=60$. For larger value of $N$, however, SSC can outperform JSC. This is due to the fact that the on-air combination discussed above may be destructive due to the lack of channel state information at the devices (see also \cite{mergen06,anandkumar07} for analogous discussions on TBMA).
\begin{figure}[h!]
\begin{minipage}[b]{1.0\linewidth}
  \centering
  \centerline{\includegraphics[width=8cm]{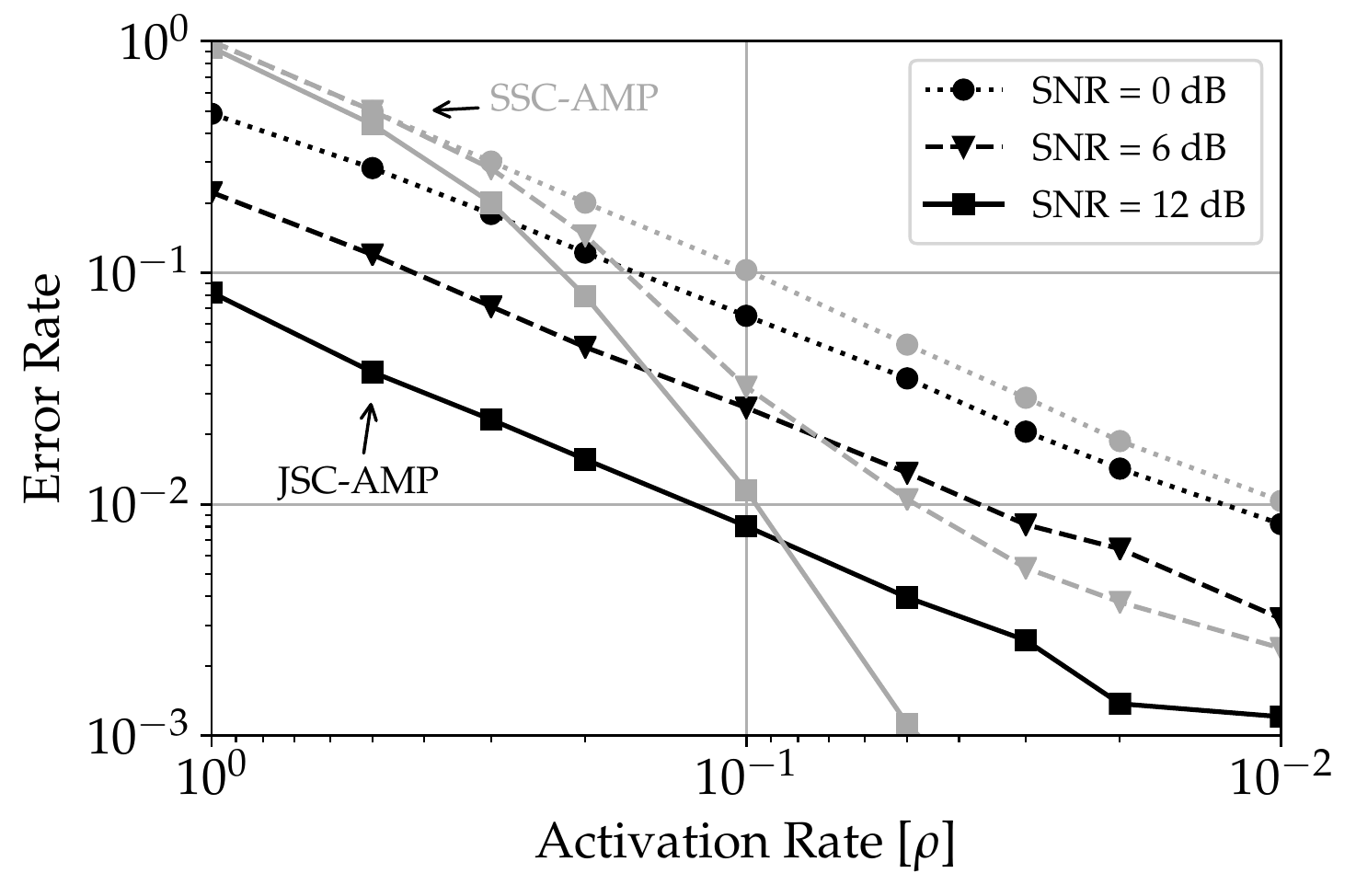}}
\end{minipage}
\caption{Decoding error-rate for SSC and JSC as function of the event activation rate $\rho$, with SNR as parameter. The number of events is $M = 24$, the group size is $G=8$, $R = 2$ and the codeword length is $N = 56$.}  
\label{fig:res_3}
\end{figure}

\vspace{-10pt}
Finally, in Fig.~\ref{fig:res_3}, we evaluate the effect of the event activation probability $\rho$ on the decoding performance of SSC and JSC for various values of the SNR. We set $G=8$ and $R=2$. We observe that JSC outperforms SSC for larger probabilities of event activation $\rho$ for all SNR values. However, as the probability of event activation $\rho$ decreases, and as the SNR increases, we observe that SSC outperforms JSC. The reason for this is again to be found in the advantages and limitations of on-air non-coherent combining enabled by JSC. As the probability of event activation decreases, SSC can leverage the increased sparsity of the vector $\boldsymbol{x}$ (as given by the SSC system model (\ref{eq:ssc_rx})) to overcome the combined effects of interference and fading. In this regime, the performance of SSC is essentially noise-limited, and the decoding performance improves significantly with the increase of the SNR. In contrast, with an increasing $\rho$ the performance of JSC is effectively limited by the fading statistics due to the potentially destructive on-air combining on shared codewords.
\vspace{-5pt}
\section{Conclusions and Future Work}
\label{sec:conclusion}
In this paper, we proposed a grant-free random access scheme that builds on joint source-channel (JSC) coding via a novel non-orthogonal variant of Type-Based Multiple Access (TBMA). The method applies to a general multi-event setting and employs  non-orthogonal codewords to improve spectral efficiency. For the proposed TBMA-based protocol, we developed a Bayesian message-passing detection scheme that jointly infers the events' activity and related scalar parameters. To evaluate the benefits of the proposed approach, we performed numerical comparison with a solution based on separate source-channel coding (SSC) under the same  Bayesian receiver processing framework. We observed that, as compared to SSC, JSC is able to operate at smaller codeword lengths, and hence at larger spectral efficiency values, by taking advantage of larger group sizes through the over-the-air combination of information from multiple devices. For future work, it would be interesting to consider a generalization of the approach to a multi-cell Fog-Radio Access Network architecture, and to assess trade-offs between edge and cloud processing \cite{Kassab19}. 
\bibliographystyle{IEEEbib}
\bibliography{refs}

\begin{thebibliography}{10}

\bibitem{polyanskiy10}
Y.~Polyanskiy, H.~V. Poor, and S.~Verdu,
\newblock ``{Channel Coding Rate in the Finite Blocklength Regime},''
\newblock {\em IEEE Transactions on Information Theory}, vol. 56, no. 5, pp.
  2307--2359, 2010.

\bibitem{Shahab19}
Muhammad {Basit Shahab}, Rana {Abbas}, Mahyar {Shirvanimoghaddam}, and Sarah~J.
  {Johnson},
\newblock ``{Grant-free Non-orthogonal Multiple Access for IoT: A Survey},''
\newblock {\em arXiv e-prints}, p. arXiv:1910.06529, Oct 2019.

\bibitem{chen14}
X.~Chen and D.~Guo,
\newblock ``{Many-Access Channels: The Gaussian Case with Random User
  Activities},''
\newblock in {\em 2014 IEEE International Symposium on Information Theory},
  June 2014, pp. 3127--3131.

\bibitem{shamai97}
S.~Shamai,
\newblock ``{A Broadcast Strategy for the Gaussian Slowly Fading Channel},''
\newblock in {\em Proceedings of IEEE International Symposium on Information
  Theory}, Jun 1997.

\bibitem{polyanskiy17}
Y.~Polyanskiy,
\newblock ``{A Perspective on Massive Random-Access},''
\newblock in {\em 2017 IEEE International Symposium on Information Theory
  (ISIT)}, June 2017, pp. 2523--2527.

\bibitem{mergen06}
G.~{Mergen} and L.~{Tong},
\newblock ``{Type Based Estimation Over Multiaccess Channels},''
\newblock {\em IEEE Transactions on Signal Processing}, vol. 54, no. 2, pp.
  613--626, Feb 2006.

\bibitem{Liu04}
{Ke Liu} and A.~M. {Sayeed},
\newblock ``{Asymptotically Optimal Decentralized Type-based Detection in
  Wireless Sensor Networks},''
\newblock in {\em 2004 IEEE International Conference on Acoustics, Speech, and
  Signal Processing}, May 2004, vol.~3, pp. iii--873.

\bibitem{anandkumar07}
A.~{Anandkumar} and L.~{Tong},
\newblock ``{Type-Based Random Access for Distributed Detection Over
  Multiaccess Fading Channels},''
\newblock {\em IEEE Transactions on Signal Processing}, vol. 55, no. 10, pp.
  5032--5043, Oct 2007.

\bibitem{Kassab19}
Rahif {Kassab}, Osvaldo {Simeone}, and Petar {Popovski},
\newblock ``{Information-Centric Grant-Free Access for IoT Fog Networks: Edge
  vs Cloud Detection and Learning},''
\newblock {\em arXiv e-prints}, p. arXiv:1907.05182, Jul 2019.

\bibitem{rangan17}
Sundeep Rangan, Alyson~K Fletcher, Vivek~K Goyal, Evan Byrne, and Philip
  Schniter,
\newblock ``{Hybrid Approximate Message Passing},''
\newblock {\em IEEE Transactions on Signal Processing}, vol. 65, no. 17, pp.
  4577--4592, 2017.

\bibitem{liu2018sparse}
Liang Liu, Erik~G Larsson, Wei Yu, Petar Popovski, Cedomir Stefanovic, and
  Elisabeth De~Carvalho,
\newblock ``{Sparse Signal Processing for Grant-free Massive connectivity: A
  Future Paradigm for Random Access Protocols in the Internet of Things},''
\newblock {\em IEEE Signal Processing Magazine}, vol. 35, no. 5, pp. 88--99,
  2018.

\end{thebibliography}
\end{document}